\documentclass[AMA,STIX2COL]{WileyNJD-v2}
\usepackage{moreverb}
\usepackage{amsmath,graphicx}
\usepackage{subfig,float}
\usepackage{multirow}
\usepackage{bm}
\usepackage{setspace}
\usepackage{rotating} 

\usepackage{color, colortbl}

\newcommand\BibTeX{{\rmfamily B\kern-.05em \textsc{i\kern-.025em b}\kern-.08em
		T\kern-.1667em\lower.7ex\hbox{E}\kern-.125emX}}

\articletype{Full paper submitted to the journal Magnetic Resonance in Medicine (MRM) }%

\received{<day> <Month>, <year>}
\revised{<day> <Month>, <year>}
\accepted{<day> <Month>, <year>}

\begin{document}
	
	\title{SNR-enhanced diffusion MRI with structure-preserving low-rank denoising in reproducing kernel Hilbert spaces}
	
	\author[1]{Gabriel Ramos-Llord{\'e}n*}
	\author[2]{Gonzalo Vegas-S{\'a}nchez-Ferrero}
	\author[3]{Congyu Liao}
	\author[2]{Carl-Fredrik Westin}
	\author[3]{Kawin Setsompop}
	\author[1,2]{Yogesh Rathi}

	\corres{*Gabriel Ramos-Llord{\'e}n. \email{gramosllorden@mgh.harvard.edu}}

	\address[1]{Department of Psychiatry, Brigham and Women's Hospital, Harvard Medical School, Boston, Massachusetts, USA}
	\address[2]{Department of Radiology, Brigham and Women's Hospital, Harvard Medical School, Boston, Massachusetts, USA}
	\address[3]{Athinoula A. Martinos Center for Biomedical Imaging, Massachusetts General Hospital, Harvard Medical School, Boston, Massachusetts, USA}
	
	\abstract[Abstract]{	{\bf Purpose:} To introduce, develop, and evaluate a novel denoising technique for diffusion MRI that leverages non-linear redundancy in the data to boost the SNR while preserving signal information. 
		{\bf Methods:} We exploit non-linear redundancy of the dMRI data by means of Kernel Principal Component Analysis (KPCA), a  non-linear generalization of PCA to reproducing kernel Hilbert spaces. By mapping the signal to a high-dimensional space, better redundancy is achieved despite nonlinearities in the data thereby enabling better denoising than linear PCA. We implement KPCA with a Gaussian kernel, with parameters automatically selected from knowledge of the noise statistics, and validate it on realistic Monte-Carlo simulations as well as with in-vivo human brain submillimeter resolution dMRI data. We demonstrate  KPCA denoising using multi-coil dMRI data also. 
		{\bf Results:} SNR improvements up to 2.7 X were obtained in real in-vivo datasets denoised with KPCA, in comparison to SNR gains of up to 1.8 X  when using state-of-the-art PCA denoising, e.g., Marchenko-Pastur PCA (MPPCA). Compared to gold-standard dataset references created from averaged data, we showed that lower normalized root mean squared error (NRMSE) was achieved with KPCA compared to MPPCA. Statistical analysis of residuals shows that only noise is removed. Improvements in the estimation of diffusion model parameters such as fractional anisotropy, mean diffusivity, and fiber orientation distribution functions (fODFs) were demonstrated. 
		{\bf Conclusion:} Non-linear redundancy of the dMRI signal can be exploited with KPCA, which allows superior noise reduction/SNR improvements than state-of-the-art PCA methods, without loss of signal information.}
	
	
	\maketitle

\section{Introduction}

Diffusion MRI (dMRI) is a noninvasive imaging modality that allows the characterization of tissue microstructure of biological tissues with an unrivaled level of quality and detail.  When diffusion-encoding gradients are played out,  the MR signal becomes sensitive to the diffusion of water molecules and their interaction with the surrounding microstructure \cite{Stejskal1965}. Hence, the dMR signal, carrying unique information, can be used to probe the microstructural environment of tissues. Unfortunately, diffusion encoding gradients as well as the $T_2$ spin-to-spin relaxation, both contribute to the characteristic attenuation of the dMRI signal during the diffusion probing time, which makes the signal-to-noise ratio (SNR) of the diffusion weighted MR images inherently low \cite{Jones2010}. This is of particular concern for high-resolution dMRI, as the SNR decreases even further due to a decrease in the voxel size. Low SNR not only complicates visual inspection but hampers quantitative analysis of informative tissue parameters, for example, by reducing the accuracy and precision of the parameter estimates \cite{Veraart2016}.

Increasing the SNR in dMRI is a major goal for the MRI community. Ultrahigh field dMRI \cite{Gallichan2018,Eichner2014,Kleinnijenhuis2015}, advanced dMRI acquisition protocols \cite{Setsompop2012,Setsompop2018,Liao2020,Haldar2020,Ramos-Llorden2020,Wu2016,Wu2016a,Eichner2020}, or noise reduction techniques \cite{Macovski1996,Eichner2015,Aja-Fernandez2008,Veraart2016}, to name a few, are complementary approaches that have been shown to enhance the SNR. In this work, we focus on noise removal techniques, in particular, the thermal noise at the reception of the Radio-Frequency signal \cite{Macovski1996}, and which further propagates to the dMRI signal domain during image reconstruction \cite{Aja-Fernandez}. 

Noise reduction or denoising can be done by averaging \cite{Eichner2015}: the subject is imaged several times with identical image parameters, and the resulting images averaged. Under some statistical assumptions of the data, this simple technique can increase the SNR by a factor of $\sqrt{N_\text{scans}}$ with $N_\text{scans}$ the number of repetitions. Evidently, this approach requires additional scan time, and as the acquisition time of a conventional dMRI protocol is already relatively long, averaging becomes impractical for routine use.

From a signal processing perspective, denoising can be seen as a post-processing approach, where an algorithm  attempts to remove the noise, i.e., reduce the noise standard deviation, while maintaining the noise-free  signal undistorted.  Trivially formulated,  denoising has been a longstanding problem in image processing, where many challenges need to be confronted,  which are further aggravated in quantitative image modalities such as diffusion MRI.  Indeed,  early computer vision-based denoising algorithms applied to dMRI have shown to be detrimental to parameter estimation quality \cite{Veraart2016}.  Some exemplar cases are the popular Total-Variation- based noise removal techniques \cite{Rudin1992,Knoll2011} or more recently, non-local means algorithms \cite{Descoteaux2008,Manjon2010}. The artefactual components that these algorithms introduce, though invisible to human eye, necessarily result in biased estimates in subsequent parameter quantification, thereby turning the improvement in noise reduction futile for quantitative analysis \cite{Veraart2016}. Other shortcomings are the loss of spatial resolution due to blurring in the presence of partial volume effects. 

On the other hand, denoising algorithms that depart from exploiting spatial similarity but leverage  ``redundancy" of the dMRI signal along diffusion direction have been shown to suppress noise significantly while preserving the dMRI signal, with no apparent blurring or biases. Principal Component Analysis (PCA) based methods belong to this category of methods. PCA-based methods were first used in   \cite{manjon2013diffusion}, and ever since, have been thoroughly validated for diffusion MRI parameter quantification and substantially refined and improved. Most of the improvements that PCA algorithms have witnessed are based on the estimation of the number of signal-only principal components. By removing the rest of the components, attributed to noise, denoising is performed. While this threshold was heuristically set in \cite{manjon2013diffusion}, the criterion was formalized in \cite{Veraart2016}, where an elegant approach was proposed relying on the theory of random covariance matrix theory, in particular, exploiting the universal Marchenko-Pastur law for eigenvalues. The MPPCA method of \cite{Veraart2016} has also been extended to general noise models other than the additive white Gaussian noise case \cite{Cordero-Grande2019}. 

All dMRI PCA denoising methods are limited by the degree of redundancy in the signal with respect to the dimensionality of the data, i.e., the number of gradient directions. Several factors, including spatial resolution, number of b-values, number of  gradient directions, determine the amount of redundancy in the data. The capability for noise reduction is hampered  if the SNR is very low as in high-resolution diffusion MRI. We would like to emphasize here that this limitation is not attributable to the PCA method at hand but in the assumption that the covariance matrix of the signal is of sufficiently low rank. 

Fortunately, there exist non-linear redundancies which are not captured by the linearity assumption implicitly adopted in PCA, but that can be exploited to enhance the SNR of dMRI substantially more than possible with current state-of-the-art approaches. The fundamental idea is to look for high dimensional non-linear spaces where the covariance matrix of the transformed dMRI signal turns out to be of low rank. Enforcing this prior knowledge, i.e., applying PCA in the transformed domain, we can denoise the signal in the high dimensional space  and map it back to the original space. As a unified approach, this operation exploits the non-linear relationships within the data, and is referred to as a non-linear generalization of PCA. The whole process can be carried out by a technique called Kernel Principal Component Analysis (KPCA) \cite{Schoelkopf1997}, where a kernel function implicitly determines the mapping to a high dimensional Hilbert space, where the dMRI signal is denoised \cite{Scholkopf1999} 

In this work, we introduce  KPCA denoising to the dMRI community, and showcase it with a Gaussian kernel that maps the dMRI signal to an infinite dimensional space where redundancy is exploited. The parameters of the kernel as well as the rank of the covariance matrix are chosen in a data-driven manner, as those that provide  the best signal representation according to the Mean Square Error (MSE) between the denoised and the underlying noise-free signal. The Stein Unbiased Risk Estimate (SURE) is used as a proxy of the MSE \cite{Stein1981,Donoho1995,Ramani2008}, circumventing the need for the unobservable noise-free signal. Only knowledge of the noise statistics is required, which we input to the algorithm with state-of-the-art noise mapping techniques.

We thoroughly validate KPCA with realistic Monte-Carlo simulations as well as with several in-vivo human brain datasets acquired with submillimeter spatial resolution. In addition, KPCA was validated with an in vivo human brain multi-coil dMRI dataset, capturing non-linear redundancies in the coil and gradient directional domains simultaneously.  In all cases, we confirmed superior noise reduction compared to the linear MPPCA method,  which immediately translates to higher SNR enhancement, while preserving signal reliably, as confirmed by residual analysis. Finally, improved diffusion parameter estimation was invariably found when dMRI datasets were denoised with KPCA. A preliminary version of this work has been presented as an abstract at the ISMRM 2020 \cite{Ramos-Llorden2020a}.

\section{Theory}
\subsection{Redundancy of dMRI signals in canonical spaces: PCA denoising  }
In dMRI, it is often assumed that the diffusion signal carries redundant information between the gradient directions (also referred to as q-space).  Redundancy can be elegantly captured by covariance matrix analysis.  Let $\bm x  \in \mathbb{R}^M$ be the diffusion MR signal with $M$ the number  gradient diffusion directions. Its second-order statistical characterization is given by its mean  $\bm{\mu} \in \mathbb{R}^N$ and covariance matrix $C_{\bm{x}} \in  \mathbb{R}^{M{\times}M}$. The diffusion signal $\bm x$ is said to be ``redundant" if $C_{\bm{x}}$ is rank-deficient, with rank $K$ substantially smaller than the dimensionality $M$. In that case, $C_{\bm{x}}$,  a low-rank matrix, can be written with eigenvalue decomposition $C_{\bm{x}} = \sum_{k=1}^K{\lambda}_k\bm{u}_k\bm{u}_k^T$.  The low-rank diffusion signal  $\bm{x}$ with these statistical properties is given by the so-called   ``spike" model
\begin{equation}
\bm{x} =  \bm \mu + \sum_{k=1}^K{\lambda}_k^{1/2}v_k\bm{u}_k\,, \label{eq:low-rank}
\end{equation} 
where $v_k$ are independent and identically distributed (IID) random variables with zero mean and unit variance. The diffusion signal is always corrupted by noise, $\bm w \in \mathbb{R}^M$, which is typically modeled as additive and with zero mean, i.e., $\bm y = \bm x + \bm w$. Denoising the signal $\bm y$ can be formulated as an estimation problem, where the goal is to estimate the noise-free signal $\bm x$ from the observed data $\bm y$.  As $C_{\bm{w}}$ is a full-rank matrix (noise is not redundant), it is precisely the low-rank nature of $C_{\bm{x}}$ in comparison to  $C_{\bm{w}}$ ($K<<M$) that allows to ``separate" signal from noise, i.e., estimating $\bm x$ reliably.  Low-rank denoising is usually performed in image patches, each one containing $N$ diffusion signals $\bm x_n$, $n=1,\ldots,N$, with mean $\bm \mu$ and covariance matrix $C_{\bm{x}}$.  For the noisy spike model,
\begin{equation}
\bm y_n =  \bm x_n +  \bm w_n = \bm \mu + \sum_{k=1}^K{\lambda}_k^{1/2}v_{kn}\bm{u}_k + \bm w_n\,, \label{eq:low-rankmodel}
\end{equation}
 the optimal estimate of $\bm x_n$ in terms of the norm  $\sum_{n=1}^N{\vert \vert \bm y_n - \hat{\bm{x}}_n \vert\vert}_2^2$ (assuming rank $K$ known)   is given by \cite{Nadakuditi2014}
\begin{equation}
\hat{\bm{x}}_n =  \hat{ \bm \mu} + \sum_{k=1}^K{\hat{\lambda}}_k^{1/2}\hat{v}_{kn}\hat{\bm{u}}_k\,, \label{eq:denoisedMRI}
\end{equation}
where $\hat{\lambda}_k$, $\hat{v}_{kn}$ and $\hat{\bm{u}}_k$ are obtained from the singular value decomposition (SVD) of  the noisy data matrix $\bm Y = [\bm y_1 - \hat{\bm \mu}, \bm y_2 - \hat{\bm \mu}, \ldots, \bm y_N - \hat{\bm \mu}]$,  that is,
\begin{equation}
\bm Y  = \sqrt{N}\hat{\bm{U}} \hat{ \bm \Lambda } ^{1/2} \hat{\bm{V}}^T 
\end{equation}
 by nullifying components with index $k>K$, and being $\hat{ \bm \mu}$ the sample mean of $\bm y_n$. This denoising framework is also called Principal Component Analysis (PCA) denoising, with $\hat{\bm{u}}_k$ as the principal components. As already mentioned in the introduction, all of the PCA-based dMRI denoising methods (\cite{Manjon2010,Veraart2016,Lemberskiy2019,Cordero-Grande2019}) fundamentally differs in the the way $K$ is estimated. No improvements are made in the intrinsic data model. In this work, we instead reconsider the low-rank model  of Eq. \ref{eq:denoisedMRI}. In particular,  we are interested in the intrinsic redundancy in the noise-free image $\bm x$.  If $K$ is comparable to $M$, the optimal solution of  Eq. \ref{eq:denoisedMRI} can hardly denoise, no matter which PCA algorithm we employ. MRI artifacts, partial volume effects, higher spatial resolution,  non-conventional dMRI sequences, to name a few, all of these factors may increase the original rank $K$ of the signal.  A possible way to increase the redundancy is to focus on $M$, rather than $K$, and increase the dimensionality of the signal by adding more gradient directions. Theoretically appealing, this \emph{modus operandi} will necessarily necessitate additional scan time, which is often of concern in clinical settings.

In the quest for data redundancy, we look for information redundancy in data domains which are not necessarily the canonical space  where conventional PCA is applied. The next section is devoted to motivate and formalize our approach, after which we present the novel denoising method in Section \ref{ss:kpca}.

\subsection{Redundancy of dMRI signals in high-dimensional Hilbert spaces}

Our starting point is a function $\bm{\phi}(\cdot)$  that maps the diffusion signal $\bm{x}$ from  the native space $\mathbb{R}^N$ to a ``feature" space $\mathcal{F}$, often high-dimensional, $P>>M$, being $P=dim(\mathcal{F})$. We will deal with the definition of  $\bm{\phi}(\cdot)$ later, but for now let us assume that the transformed  data $\bm{\phi}( \bm{x} )$  is redundant, i.e., the covariance matrix of $\bm{\phi}( \bm{x} )$ is of rank $K_\mathcal{F}<< P$,  even if native rank $K$ is high. For example, any mapping which makes data ``sparser" in the feature domain will ``compress'' the information more than in the native space. This redundancy  translates into a low-rank covariance matrix $C_{\bm{\phi}( \bm{x} )}$. We can exploit this redundancy to denoise data in the feature space and eventually return to the native space to get the denoised signals $\bm{\hat{x}}_n$. Note that the entire process can be seen as a way to exploit non-linear redundancy that the dMRI signal can carry, and that could be otherwise difficult to capture with conventional PCA.

In  $\mathcal{F}$, similar to PCA, the optimal estimates of $\bm \phi(\bm x_n)$ of rank $K_\mathcal{F}$ are those  minimizing the  error $\sum_{n=1}^N{\vert \vert \bm \phi (\bm{y}_n) - \hat{\bm \phi}(\bm x_n)\vert\vert}_2^2$. They can be shown to be the projection of the mapped noisy signals $\bm{\phi}(\bm{y}_n)$, $n=1,\ldots,N$, onto the feature space, $P{\bm{\phi}}(\bm{y}_n)$, 
\begin{equation}
\hat{\bm \phi}(\bm x_n) = P{\bm{\phi}}(\bm{y}_n) \triangleq  \bm {\bar{\phi}} + \sum_{k=1}^{K_\mathcal{F}}{\hat{\lambda}}_k^{1/2}\hat{v}_{kn}\hat{\bm{u}}_k \,.\label{eq:denoisedMRIfeature}
\end{equation}
In Eq. \ref{eq:denoisedMRIfeature}, $\bm {\bar{\phi}}$ is the mean of  $\bm{\phi}(\bm{y}_n)$, $n=1,\ldots,N$, $\hat{\bm{u}}_k$ are the non-linear principal component directions, and the rest of parameters are obtained (we maintain the notation of Eq. \ref{eq:denoisedMRI}) from the SVD of centered noisy projected data matrix 
\begin{equation}
\bm \Phi = [\bm{\phi}( \bm{y}_1 ) - \bm {\bar{\phi}}  , \bm{\phi}( \bm{y}_2 ) - \bm {\bar{\phi}} , \ldots,\bm{\phi}( \bm{y}_N ) - \bm {\bar{\phi}}  ] \,.
\end{equation}
 While the low rank denoising is performed in $\mathcal{F}$, we would like to come back to the native space.  If we want to denoise the signal $\bm y^*$ at the center of the patch,  we then look for that $\bm{x}$, which after being mapped to the feature space, $\bm \phi(\bm x)$, turns out to be the closest to  the projection $P{\bm{\phi}}(\bm{y}^*)$ (Eq. \ref{eq:denoisedMRIfeature}), i.e.,
\begin{equation}
\hat {\bm {x}}^*= \operatorname*{arg\,min}_{{\bm{x}}} {\vert \vert  \bm{\phi}( \bm{x} ) -  P{\bm{\phi}}(\bm{y}^*)\vert \vert}_2^2 \,.
\label{eq:l2norm}
\end{equation}

A fundamental result that is of high relevance for this work is the following.  To apply PCA in the feature space $\mathcal{F}$ defined by the mapping $\bm{\phi}(\cdot)$, and to solve Eq. \ref{eq:l2norm} in order to return to the native space, we do not need to know $\bm{\phi}(\cdot)$ explicitly, but just the inner product of the form $\langle \bm{\phi}(\bm x),\bm{\phi}(\bm y) \rangle$ for $\bm x$ and $\bm y$ in $\mathbb{R}^M$.  Since $\langle \bm{\phi}(\cdot ),\bm{\phi}(\cdot ) \rangle$ is a symmetric, positive definite function, it automatically defines a kernel function in $\mathbb{R}^M \times \mathbb{R}^M$ as $k(\bm x, \bm y ) =  \langle \bm{\phi}(\bm x),\bm{\phi}(\bm y) \rangle$. Conversely, choosing a kernel function $k(\cdot,\cdot)$ implicitly defines a mapping $\bm \phi( \cdot)$ \cite{Scholkopf1999}. Therefore, it is the kernel function that implicitly defines the feature space. Features spaces with this property are called reproducing kernel Hilbert spaces, and applying PCA in the feature space is termed Kernel Principal Component Analysis \cite{Scholkopf1999}.  In the next section, we present our Kernel PCA denoising method in detail, elaborating on the selection of the kernel as well as the rank  $K_\mathcal{F}$.

\section{Methods}
\subsection{Kernel Principal Component Analysis (KPCA) denoising}
\subsubsection{KPCA algorithm}
\label{ss:kpca}
Given a kernel $k(\cdot,\cdot)$, the denoised signal $\hat {\bm {x}}^*$ in the feature space defined by $k(\cdot,\cdot)$ can be written as \cite{Mika1999}
\begin{equation}
\hat {\bm {x}}^ * = \operatorname*{arg\,min}_{{\bm{x}}}   k({\bm{x}},{\bm{x}}) - 2\sum_{n=1}^N{\gamma}_nk({\bm{x}},{\bm{y}}_n) \,,
\label{eq:l2kernel}
\end{equation}
with $\bm \gamma = {[\gamma_1, \gamma_2, ..., \gamma_N ]}^T = \sum_{k=1}^{K_\mathcal{F}}\beta_k\bm\alpha_{k} +1/N(1-\bm 1^T\sum_{k=1}^{K_\mathcal{F}}\beta_k\bm\alpha_{k})$, $\bm 1$ an $N$-dimensional column vector with all entries equal to one, and $\bm\alpha_{k}$, the first $K_\mathcal{F}$ eigenvectors that solve the following eigenvalue problem
\begin{equation}
\bm H \bm K \bm H\bm \alpha_k  = N\hat{\lambda}_k\bm \alpha_k \quad  \text{with} \quad  {N\hat{\lambda}_k} {\vert\vert  \bm {\alpha}_k \vert \vert}_2^2 = 1 \,, \label{eq:eigenkpca}
\end{equation}
where $\bm H = \bm I - \frac{1}{N}\bm 1 \bm 1^T$ is a ``center" matrix and  $\bm K$,  the so-called kernel matrix ($N \times N$), $K_{mn}= k({\bm{y}}_n,{\bm{y}}_m) $.  Finally, the coefficients $\beta_k$, the components of projection of $\bm \phi(\bm {y^}*)$ onto the k-th non-linear principal component $\hat{\bm{u}}_k$, can be  computed as 
\begin{equation}
\beta_k = \sum_{n=1}^{N}\alpha_{kn}\tilde{k}(\bm {y^}*, \bm y_n) \,,
\end{equation}
with $\alpha_{kn} $ the $n$-th coefficient of ${\bm \alpha}_k$, and $\tilde{k}(\bm {y^}*, \bm y_n)$ equal to \cite{Kwok2004}
\begin{align}
\tilde{k}(\bm {y^}*, \bm y_n) & =  k(\bm {y^}*, \bm y_n) - \frac{1}{N}\sum_{i=1}^Nk(\bm {y^}*, \bm y_i) \nonumber \\
 & - \frac{1}{N}\sum_{i=1}^Nk(\bm y_i , \bm y_n) +  \frac{1}{N^2}\sum_{i,j=1}^Nk(\bm y_i, \bm y_j)  \,.
\end{align}
The interested reader can find the mathematical proof of the derivation of KPCA in \cite{Mika1999} and in Section 1.1 of the supplementary file of this submission
\subsubsection{The choice of the kernel function}
\label{ss:kernel}
We showcase KPCA denoising for dMRI with a Gaussian kernel function, 
\begin{equation}
k(\bm y_i, \bm y_n) = e^{-\frac{  {\vert \vert \bm y_i - \bm y_n \vert \vert}_2^2 } {2 h^2} } \,,
\end{equation}
with $h$ the scale parameter. Gaussian kernels have shown excellent performance in machine learning tasks and are particularly interesting for dMRI denoising for the following reasons. The implicit feature space  that the Gaussian kernel function generates can be shown to be infinite-dimensional \cite{Scholkopf1999}. As data tend to be sparser in high-dimensional spaces, higher redundancy is achieved by mapping the data with ${\bm \phi}_{h}(\cdot)$. As implied by the notation, we can control the shape of the mapping with the scale parameter $h$, and, in fact, the components of $\bm {\phi}_h (\bm y_n)$ decay with  increasing $h$.  In that sense, by varying $h$, we can adapt the level of redundancy of the dMRI signal in the feature space. This aspect will be of high interest for the automatic selection of parameters. Finally, it is possible to demonstrate that, when $ h \to \infty$,  KPCA with a Gaussian kernel behaves as linear PCA in the canonical space\cite{Jorgensen2011}.   Hence, being linear PCA then a particular case of KPCA with Gaussian kernel functions,  it is expected that our KPCA denoising  will perform typically better, as we confirm in this paper. More details about the implicit mapping related to the Gaussian kernel and the demonstration of the asymptotic equivalence of KPCA and PCA are given in the supplementary file, (Sections 1.2 and 1.3, respectively).

In addition, there are computational advantages  in choosing the Gaussian kernel. The solution of Eq. \ref{eq:l2kernel} can be obtained in very short computational time with the approximation  given in  \cite{Rathi2006}
\begin{align}
\hat {\bm {x}}^* & = \frac{   \sum_{n=1}^N{\gamma}_n \operatorname{exp} \left( -\frac{  {\vert \vert \hat {\bm {x}}^*  - \bm y_n \vert \vert}_2^2 } {2 h^2}  \right)  \bm{y}_n   }            {   \sum_{n=1}^N{\gamma}_n \operatorname{exp} \left( -\frac{  {\vert \vert \hat {\bm {x}}^*  - \bm y_n \vert \vert}_2^2 } {2 h^2}  \right)   } \nonumber \\
& \approx  \frac{   \sum_{n=1}^N{\gamma}_n \left(1- 1/2  {\vert \vert   P{\bm{\phi}}(\bm{y}^*)    - \bm \phi(\bm y_n) \vert \vert}_2^2 \right) \bm y_n }            {  \sum_{n=1}^N{\gamma}_n\left(1- 1/2  {\vert \vert P{\bm{\phi}}(\bm{y}^*) - \bm \phi(\bm y_n) \vert \vert}_2^2 \right)  }  \,.
\end{align}
with ${\vert \vert P{\bm{\phi}}(\bm{y}^*) - \bm \phi(\bm y_n) \vert \vert}_2^2$ calculated analytically. Details are given in Section 1.4 of supplementary file. Finally, in the discussion session, we elaborate on possible improvements of KPCA denoising by selecting more complex kernels.

\subsubsection{Automatic parameter selection driven by  noise statistics}
Two parameters need to be selected for our KPCA method: the scale parameter $h$  and the rank $K_\mathcal{F}$. Ideally, we would like to select those that best represent the noise-free signal $\bm x^*$, for example, by quantifying the Mean Squared Error, (risk) $E\{ {\vert \vert  \bm x^* -\hat {\bm {x}}^*(h,K_\mathcal{F}) \vert \vert }_2^2\}$ for different choices of $h$ and $K_\mathcal{F}$.  Obviously,  the ground-truth signal  $\bm x^*$ is unobservable, and hence MSE is not computable. Instead, we use the Stein Unbiased Risk Estimate (SURE) \cite{Stein1981}. Minimizing SURE can act as a surrogate for minimizing the MSE, with the critical difference that it does not require knowledge of $\bm x^*$ \cite{Donoho1995}. For an AWGN model like  that of Eq. \ref{eq:low-rankmodel}, SURE can be computed from the noisy signals $\bm y_n$, the denoised signal $\hat {\bm {x}}^*(h,K_\mathcal{F})$, and the standard deviation of the noise, $\sigma$.  We estimate the noise maps of the DWI images using the method presented in \cite{aja2015spatially}, with the assumption of Gaussian distributed data, which holds in our experiments as we show in the subsequent section. Then, for every voxel in the image patches, we fix  $\sigma$ and applied grid search minimization to get the optimal $h$ and $K_\mathcal{F}$ w.r.t. the SURE cost-function.   We used the efficient implementation of the SURE method  based on Monte-Carlo sampling \cite{Ramani2008}. We refer the reader to Section 1.5 of the supplementary file for more details about the SURE method for optimal parameter selection. In addition, a discussion is provided at the end of the paper about the extension of SURE to other noise models as well as different techniques to estimate $h$ and $K_\mathcal{F}$ that may be of interest. 

\begin{figure*}[h]
	\centering
	\includegraphics{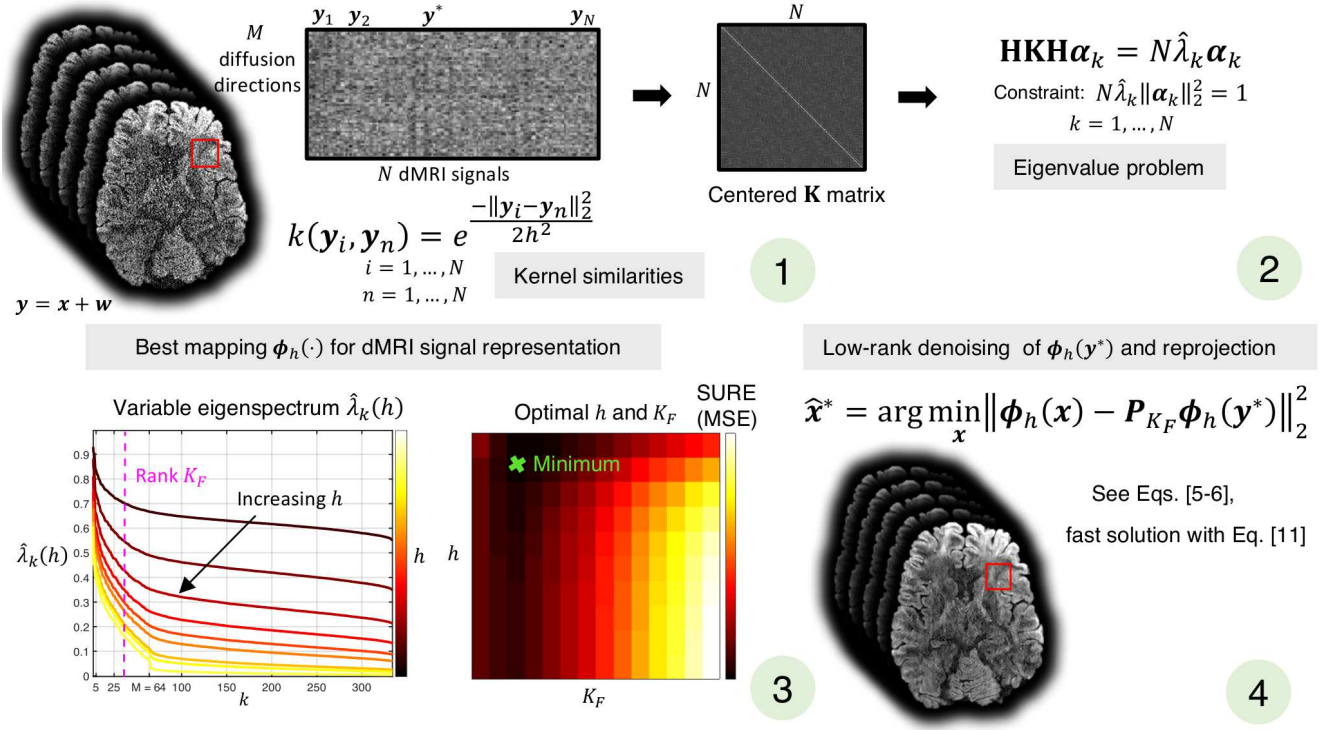} 
	\caption{Sliding-window version of KPCA denoising. For a given patch of $N$ dMRI noisy signals, (1) the similarity between each pair, $\bm y_i$, $\bm y_n$,  are calculated as $k(\bm y_i, \bm y_n)$, and represented by the centered $K$ matrix, $\bm H \bm K \bm H$. Solving the eigenvalue problem (2), we obtain the eigenvectors $\bm \alpha_k$ and eigenvalues $\hat{ \lambda}_k$ associated to the mapping $\bm {\phi} _h(\cdot)$, which defines the feature space where linear PCA is performed.  The best feature space, i.e, optimal $h$, and optimal rank $K_\mathcal{F}$ are selected as those that, after reprojection of the low-rank denoised signal to the native space,  give the best signal representation in terms of the Stein Unbiased Risk Estimate (SURE)  (3). Finally, the denoising signal at the center of the patch (4) is obtained by applying low rank denoising (with optimal $K_\mathcal{F}$)  in the optimal feature space  and reprojecting.}
	\label{fig:Figmethod}
\end{figure*}

An illustrative scheme of the KPCA denoising method used in this work is presented in Fig. \ref{fig:Figmethod}.

\subsection{Experimental validation }
We validated KPCA denoising  using simulated and in-vivo human brain dMRI data, both quantitatively and qualitatively, and compared our results with MPPCA denoising \cite{Veraart2016}. Both algorithms were implemented in a sliding-window fashion, where for each patch, only the signal at the center was denoised. The selection of the parameters for KPCA was done as follows. The standard deviation of the noise was estimated with the method of \cite{aja2015spatially}. The set of possible values of $K_\mathcal{F}$ to minimize the SURE was chosen to be in the range $[1,30]$, and the scale parameter of the Gaussian kernel $h$, was parameterized by $ h = c \; \sigma_{\text{min-class}}$ where $\sigma_{\text{min-class}}$ is the average minimum distance between all pairs of signals in the patch \cite{Rathi2006}, and the values for  $c$ was chosen from ten equidistant points in the interval $[0.6,6]$ (see Section 1.4 of the supplementary file).

\subsubsection{Simulations}
A Monte-Carlo based experiment was conducted to assess the benefits of KPCA denoising in subsequent diffusion parameter estimation. Similar to the patch-based simulation experiment in \cite{Veraart2016}, we generated $5 \times 5 \times 5 $ signals based on a diffusion tensor (axially symmetric) model and a total of $M$ gradient directions uniformly spread on the sphere with a given b-value, $b$. The underlying fractional anisotropy (FA) and mean diffusivity (MD) for each tensor in the pach was sampled from a distribution with fixed mean $\text{FA}_\text{GT}$ and $\text{MD}_\text{GT}$, and a standard deviation of 10\% with respect to the mean. 
$\text{MC} = 5000$  zero-mean uncorrelated Gaussian noise realizations were added to each of the  noise-free $N=125$ signals. The standard deviation of the noise was parameterized by a nominal SNR value, i.e.,  $\sigma = 1/\text{SNR}$. 
 
The $\text{MC} = 5000$ noisy patches were then denoised with MPPCA and KPCA, and the denoised signals were compared to the ground-truth signal. Experiments were conduced for different  a) number of diffusion directions, $M \in [32,64,128]$,  b) b-values, $b \in [1200, 1500, 2500 ]$ $s/\text{mm}^2$, c) SNR values, SNR $\in [5, 8, 15]$, and d) representative FA in both gray and white matter, $\text{FA}_\text{GT}  \in [0.2, 0.6] $. $\text{MD}_\text{GT} = 8\cdot 10^{-4} \text{ mm}^2 /s$  was considered in both cases.

The Normalized Root Mean Square Error (NRMSE) was used to compare the denoised signals with respect to the  ground-truth signal.   Diffusion tensor parameters were estimated from the log-linearized signals with a Linear Least Squares (LLS) estimator. Next,  the FA and MD were estimated and compared to the ground-truth FA and MD, $\text{FA}_\text{GT}$ and $\text{MD}_\text{GT}$.

\subsubsection{In vivo human brain submillimeter resolution dMRI data}
Whole human brain in-vivo submillimeter dMRI data were acquired and reconstructed with the generalized slice dithered enhanced resolution (gSlider) technique \cite{Setsompop2018}, and denoised with KPCA and MPPCA.  Two datasets with different spatial resolutions were considered.

\paragraph{660 ${\mu}m$ isotropic gSlider data\\}
A total of 46 thick sagittal slices were acquired (Siemens 3T Connectom scanner) with in-plane resolution 660 ${\mu}m$ and matrix size $332 \times 180$, covering the full brain (FOV = $220 \times 118 \times 151.8 \text{ mm}^3$).  The diffusion protocol consisted of $M$ = 64 diffusion-weighted images (diffusion directions uniformly distributed along sphere) with $b$ = 1500 $s/{\text{mm}}^2$ and 7 b0-images. Data was acquired \cite{Haldar2020} with a single-shot EPI sequence: Muti-Band = 2, partial Fourier = 6/8, phase-encoding (superior-inferior axis) under-sampling factor $R_{\text{in-plane}} = 2$, TR/TE = 4400/80 ms, five radio-frequency encoding pulses. The total acquisition time was about 25 min. Three repetitions were acquired to construct a gold-standard reference. 
Conventional gSlider \cite{Setsompop2018} was used to reconstruct the data and obtain whol- brain isotropic 660 ${\mu}m$ resolution. Prior to gSlider reconstruction, slice and in-plane GRAPPA was used for k-space and SMS reconstruction, and real-valued data was obtained with background phase correction \cite{Eichner2015}. Eddy-current and motion were corrected between all acquisitions using the FSL technique \texttt{FLIRT}.  The three datasets were then denoised with KPCA and MPPCA, and compared to the averaged dataset, which is considered here as the gold-standard reference.  Both algorithms were implemented in a voxel-wise fashion, with a sliding window of ${ [ 5 \times 5 \times 5] }$ voxels.

\paragraph{860 ${\mu}m$ isotropic gSlider data\\}
Whole human brain gSlider-SMS data were collected from a healthy male volunteer on a Siemens 3T Prisma scanner. Four scans of the full brain (FOV = $220 \times 220\times 163 \text{ mm}^3$)  were obtained.
A total of 38 thick axial slices were acquired with in-plane resolution of 860 ${\mu}m$ and matrix size $256 \times 256$. The diffusion protocol consisted of 64 diffusion-weighted images (diffusion directions uniformly distributed along sphere) at $b$ = 2000 $s/{\text{mm}}^2$ and 8 b0-images. Data was acquired \cite{Ramos-Llorden2020} with a single-shot EPI sequence: Muti-Band = 2, partial Fourier = 6/8, phase-encoding (posterior-anterior axis) under-sampling factor $R_{\text{in-plane}}  = 3$, TR/TE = 3500/81 ms, five radio-frequency encoding pulses. The total acquisition time was about 20 min. Four repetitions were acquired to construct a gold-standard reference. 
Data was preprocessed and reconstructed as described for the 660 ${ \mu}m$ case. After affine registration, one of the datasets was denoised with KPCA and MPPCA (identical window size as before), and compared to the averaged dataset, the gold-standard reference.

\paragraph{Quantitative validation\\}

We assessed the performance of KPCA denoising in signal preservation and parameter estimation, quantitatively. The NRMSE was used to compare the signal of the denoised diffusion-weighted  (DW) images with the signal of the averaged data set. To assess the ability of KPCA denoising for SNR enhancement, we estimated the noise maps (noise standard deviation) of the denoised datasets with the  homomorphic approach \cite{aja2015spatially}.  The SNR gain was defined as the ratio between the standard deviation of the noise in the original dataset and that of the denoised datasets. To demonstrate that KPCA preserves the underlying diffusion signal reliably, we calculated the normalized residuals between the noisy datasets and the denoised versions, and checked if any anatomical structure was present \cite{Veraart2016}. 

We conducted  DTI analysis and High Angular Resolution Diffusion Imaging (HARDI) validation. FA and MD maps were estimated with \texttt{dtifit} from FSL, and compared to the maps from the reference set. The NRMSE was used to assess the improved quality in parameter estimation. HARDI analysis was carried out with MRtrix3 \cite{Tournier2019}. Fiber Orientation Distribution Functions (fODFs) were calculated in white matter area only, with the single-shell single-tissue Constrained Spherical Deconvolution (CSD) technique \cite{Tournier2007}. For each voxel, main fiber peaks were extracted, and the angular error compared to those from the reference set were calculated. The variability in the estimation of the ODF peaks was probed with the coherence metric, $\kappa$, ($\kappa \in [0,1]$), which was originally proposed in \cite{Jones2003} and used in \cite{Veraart2016}. A high value of $\kappa$ indicates low angular variability, that is, high angular precision.

\subsubsection{Capturing non-linear coil and diffusion redundancy simultaneously} 
We investigated whether KPCA denoising can work at the reconstruction level, for example, by denoising multi-coil and diffusion data simultaneously. To that end,  we used an in-vivo human brain DW image data set comprising of one b0-image and 15 diffusion gradient directions that were uniformly spread over the sphere  (b = 1200 $s/\text{mm}^2$). The acquisition protocol was as follows. With a 3T Philips scanner, an axial slice was acquired with a single-shot EPI sequence, matrix size = $70 \times 91$, in-plane resolution of 2 mm, multi-coil system with eight channels and no-undersampling factor. To create a gold-standard reference, 20 repetitions of the same axial slice were obtained. Prior to denoising, k-space data was transformed to image space with an inverse Fourier transform.  Phase estimation was obtained by taking the complex argument of the image resulting from the inverse Fourier transform of low-pass filtered k-space data (center of the k-space).  Complex conjugate phase correction was applied, and the real part was retained. No statistical correlation was assumed between the different coils.

To apply denoising, the coil and the diffusion dimension were merged into a single dimension, $M= 8$ channels $\times$ 15 diffusion directions = 120. The size of the patch was $[7 \times 7] $. As in the previous experiment, the NRMSE, the noise maps, and the normalized residuals were calculated both from the denoised dataset with MPPCA and KPCA. NRMSE maps for diffusion derived metrics, FA and MD were also computed.

\section{Results}
\subsection{Simulations}
NMRSE results for the case b = 1200 $s/\text{mm}^2$ and $M=64$ direction, are shown as bar plots in Figure \ref{fig:Figsimul}, whereas the rest of results are shown in Table format in the supplementary file for brevity  (Table S1, S2 and S3). In general, KPCA achieves the lowest NRMSE results in signal quality, FA and MD, and shows particularly drastic improvement in FA for low nominal SNR values, e.g., SNR = 5 and SNR = 8, which correspond to real SNR values (defined as the the noise-free dMRI signal divided by sigma) of 2 and 3, typically encountered in real data noisy scenarios as those like submillimeter resolution data shown in this paper. Interestingly, the superiority of  KPCA over MPPCA denoising becomes more notorious for reduced number of diffusion directions $M= 32 $ or $M=64$. This is attributed to the lack of enough ``linear"  redundancy of the dMRI signal $K$ compared to low values of dimensionality $M$. KPCA, however, as it performs low-rank denoising in a high-dimensional space ($P>>M$), can achieved superior noise reduction, and, hence, improved parameter estimation. 

\begin{figure*}[t]
	\centering
	\includegraphics{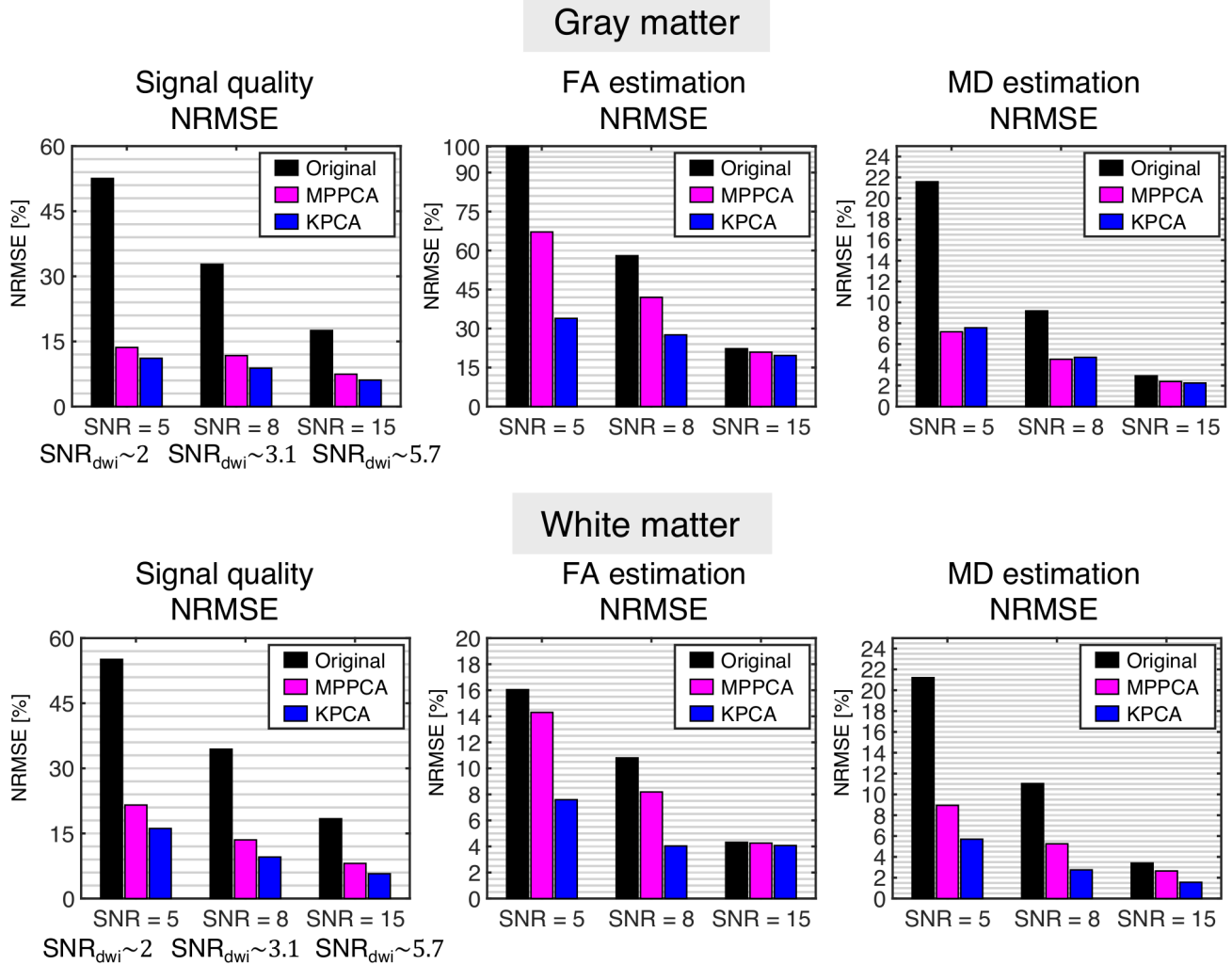} 
	\caption{Quantitative results from the MC-based simulation experiment. For representative cases of both white and gray matter, $b = 1200$ $s/\text{mm}^2$ and $M = 64$, the NRMSE [\%] of the dMRI signal, FA and MD estimates are shown for different SNR values (the corresponding mean SNR over all diffusion directions, $\text{SNR}_\text{dwi}$, is given as a reference)}
	\label{fig:Figsimul}
\end{figure*}

\subsection{In vivo human brain submillimeter resolution dMRI data}

\begin{figure*}[t]
	\centering
	\includegraphics{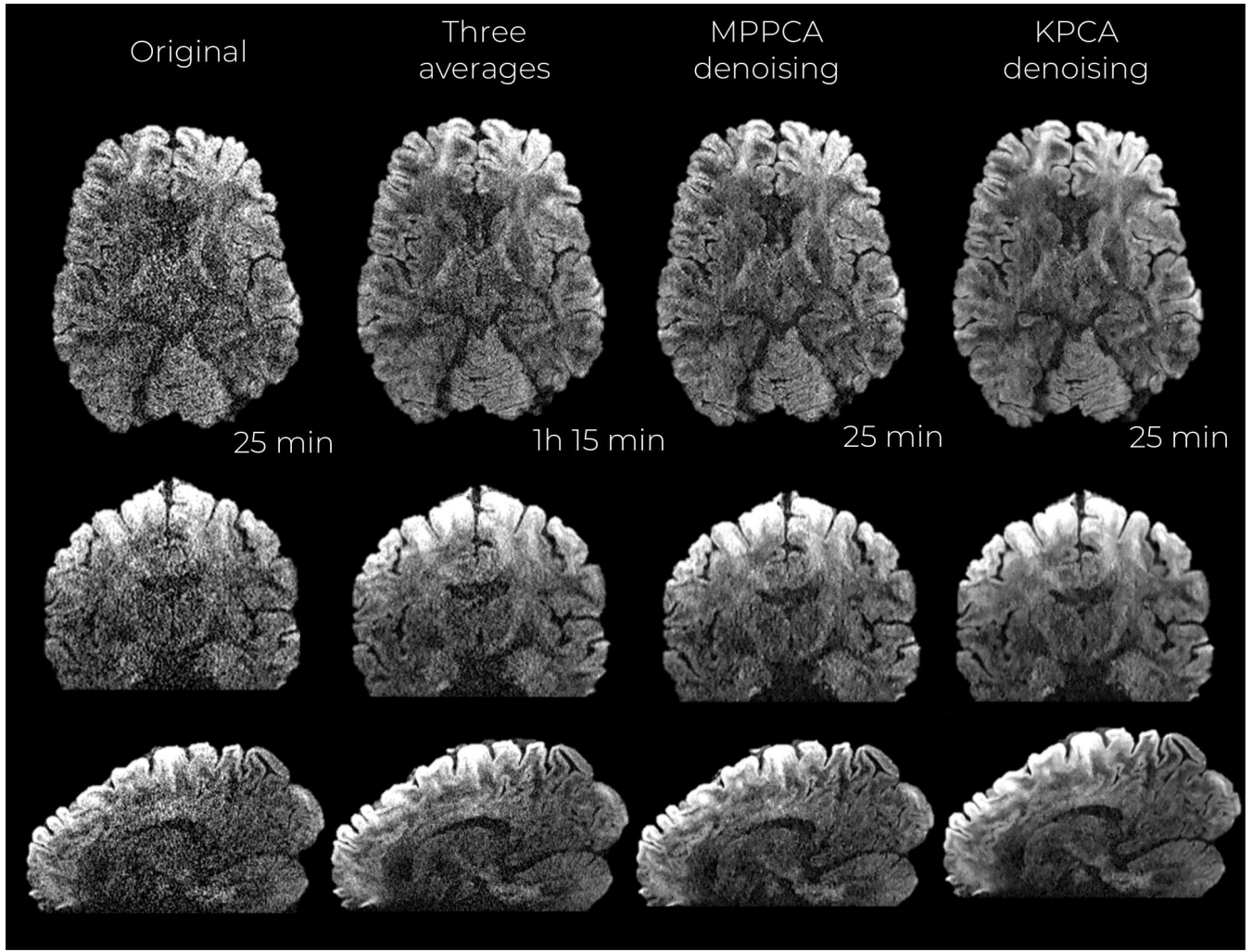} 
	\caption{Mid-axial, coronal and sagittal slices of  denoised DW images at 660 ${\mu}m$ isotropic resolution and b-value of $b $ = 1500 $s/{\text{mm}}^2$. Acquisition times are reported as well. }\label{fig:DWI660}
\end{figure*}

\begin{figure*}[t]
	\centering
	\includegraphics{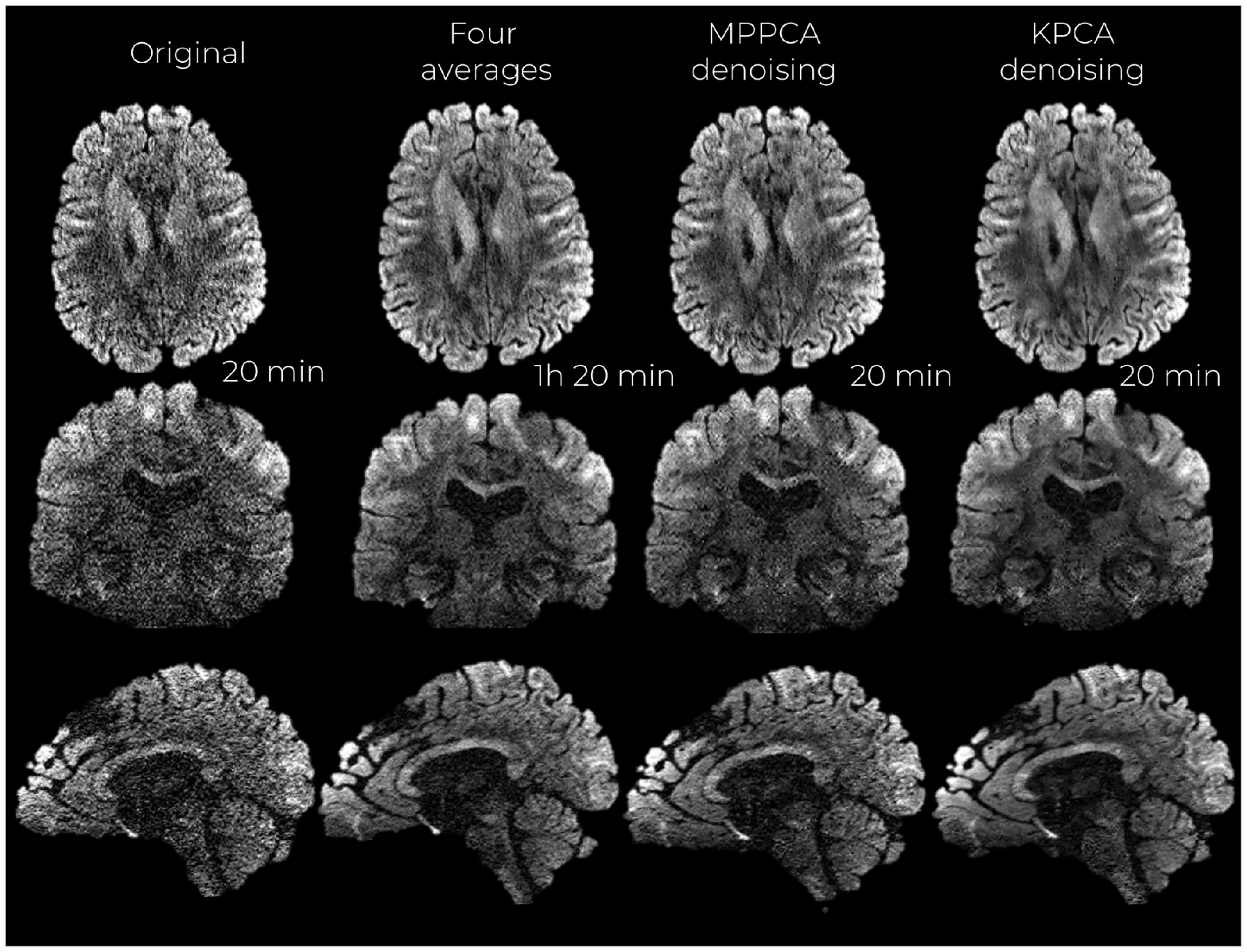} 
	\caption{Mid-axial, coronal and sagittal slices of  denoised DW images at 860 ${\mu}m$ isotropic resolution and b-value of $b $ = 2000 $s/{\text{mm}}^2$. Acquisition times are reported as well.}\label{fig:DWI860}
\end{figure*}

Denoised images with MPPCA and KPCA from the 660  and 860 $\mu$m resolution datasets are shown in Fig.\ref{fig:DWI660} and Fig.\ref{fig:DWI860}, respectively. The original dataset (no denoising) as well as the gold-standard reference, three averages for the 660 $\mu$m case and four averages for  860 $\mu$m case are also shown. 

Visually, KPCA denoising achieves a higher noise suppression than MPPCA without signal loss. No anatomical structure can be seen in the `residual' images (original - denoised DW image)  that are shown in Fig. \ref{fig:Residuals660} and Fig. S2 of the supplementary file (860 $\mu$m dataset). 

\begin{figure}[h]
	\centering
	\includegraphics{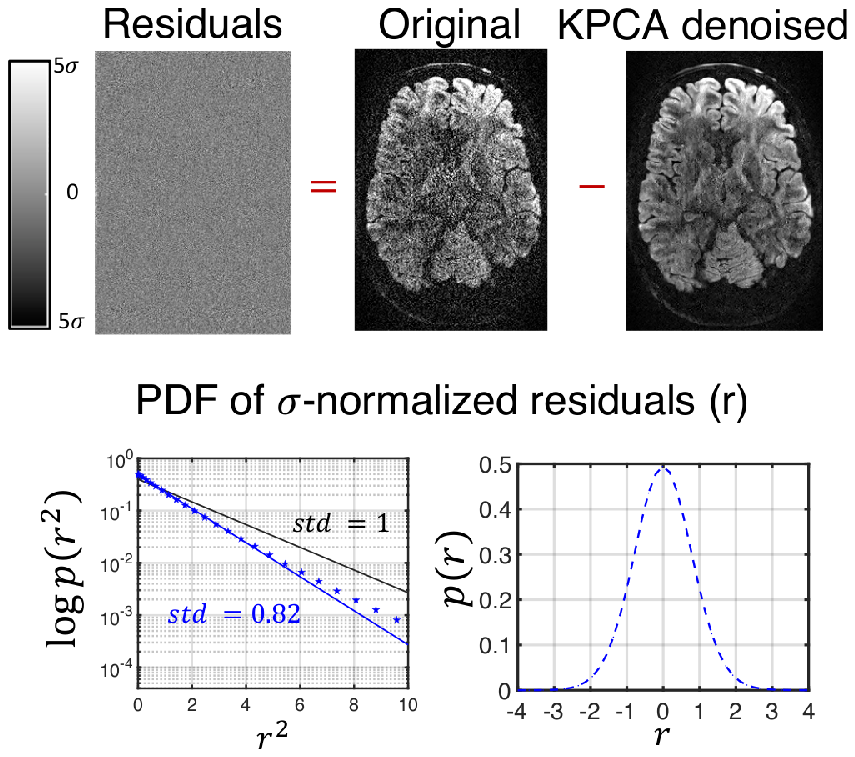} 
	\caption{Residual maps from the 660 ${\mu}m$ resolution datasets after being denoised with KPCA. On top of the figure, the residual map from a given DW image, which shows no anatomical information.   On the bottom, the probability density function of the residuals ($r$)  normalized by the level of noise, $\sigma$. For the statistics, the normalized residuals are taken for all diffusion directions and number of repetitions. Note that the residuals for KPCA approximately follows a Gaussian distribution (blue dotted line on both plots representing the estimated pdf). On blue solid-line the optimal analytical zero-mean Gaussian distribution that best fits the data (Maximum Likelihood sense). Note that the standard deviation of the normalized residual, 0.82, is lower than 1 (black-line represents a zero-mean standard Gaussian distribution).
	}\label{fig:Residuals660}
\end{figure}

Statistical analysis of residuals confirms signal preservation in KPCA denoising. Any anatomical structure in the residual dataset will make the standard deviation higher than the noise standard deviation, $\sigma$ \cite{Veraart2016}.  By analyzing the $\sigma$ -normalized residuals $r$ \cite{Veraart2016}, we found that in both cases, 660 and 860 ${\mu}m$ resolution data,  $r$ approximately follows a zero-mean Gaussian distribution with standard deviation 0.82 and 0.79, respectively, see blue dotted-blue line graphs representing the estimated pdf, $p(r)$, of normalized residuals (logarithmic plot on the left, linear plot on the right). As the standard deviation is lower than the unit (see solid black-line representing zero-mean standard Gaussian distribution),  we then can conclude that no anatomical structure is lost in KPCA denoising. The pdf $p(r)$ was estimated with a kernel density estimator. The solid blue-line represents the analytical Gaussian pdf that best fits the data in a Maximum Likelihood sense.

The estimated noise maps after denoising are presented in  Fig. \ref{fig:NRMSE660} and Fig. S3 of the supplementary material.  Note that the noise mapping method \cite{aja2015spatially} we use to estimate $\sigma$ assumes either a Gaussian or Rician distribution. Since, by assumption, the original data is Gaussian distributed (real-value phase corrected images \cite{Eichner2015}) and the residuals are shown to be Gaussian, our assumptions are well founded.

\begin{figure*}[h]
	\centering
	\includegraphics{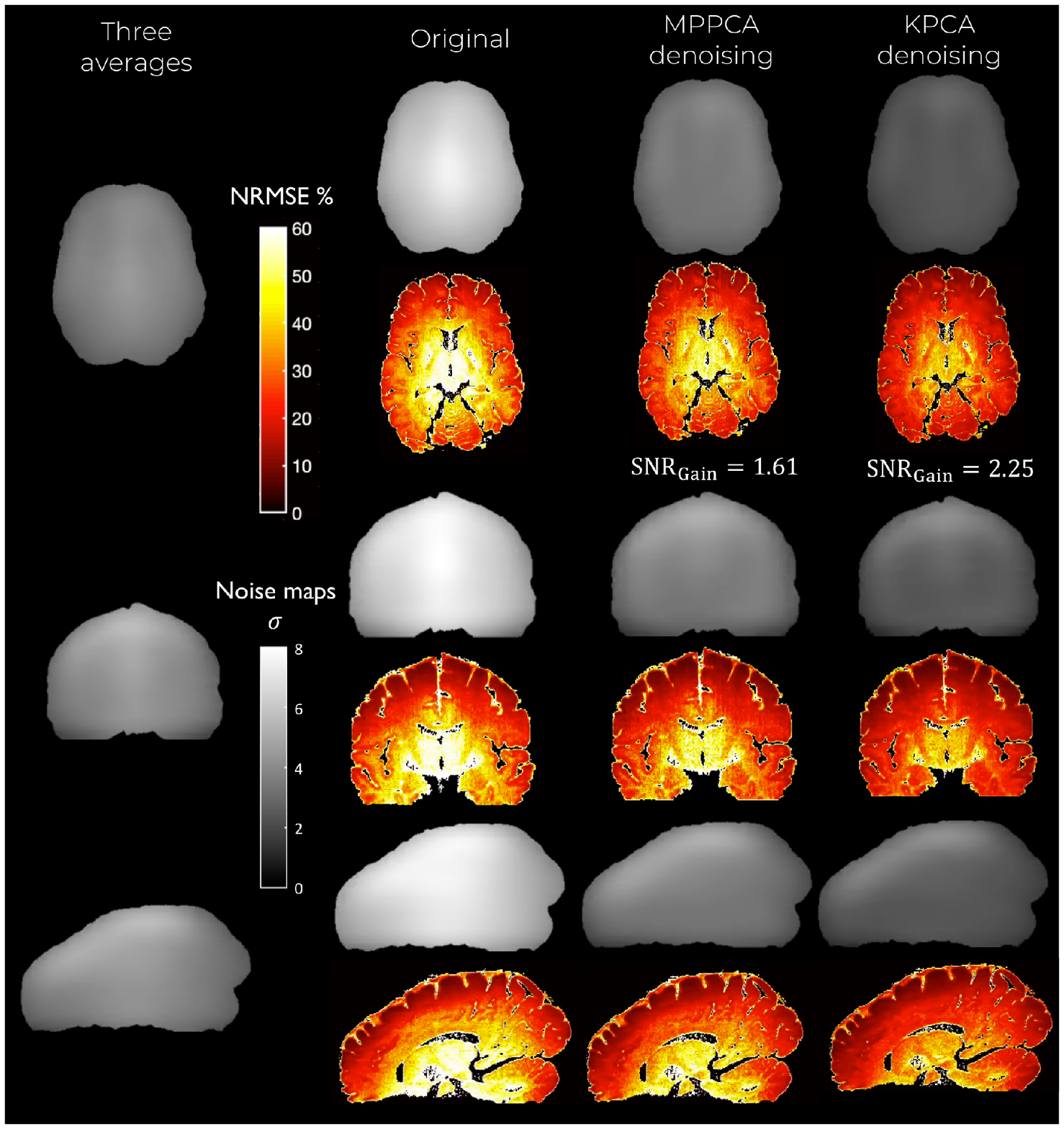} 
	\caption{Maps of the NRMSE (hot colormap) and noise level (gray colormap) for the denoised DW images at 660 ${\mu}m$ isotropic resolution and b-value of $b $ = 1500 $s/{\text{mm}}^2$. Observe that KPCA denoising obtains the lowest level of noise (highest SNR gains) and lowest NRMSE.} \label{fig:NRMSE660}
\end{figure*}

KPCA achieves higher noise suppression while still reliably preserving signal, supported by the previous experiment with residuals. The lowest levels of noise are found when the original data is denoised with KPCA, indicating that KPCA enhances the SNR to a greater extent than what is achievable  with MPPCA denoising. The SNR gain is more than 60\% higher than those obtained with MPPCA, see Table. \ref{tab:table_new}. Superior noise removal performance as well as reliable signal preservation make the NMRSE (compared to the averaged data case) substantially lower than MPPCA, both in white and gray matter (Table. \ref{tab:table_new}).  

\begin{table*}[h]
	\centering
	\begin{center}
			\begin{tabular}{l|l|l|l| l|l|l| l|l|l|  l|l|l|  l|l|l| l|l|l|}
				\cline{2-19}
				& \multicolumn{3}{|c|} {Signal} & \multicolumn{3}{|c|} {SNR} & \multicolumn{3}{|c|} {FA} & \multicolumn{3}{|c|} {MD}  & \multicolumn{3}{|c|} {Angular} & \multicolumn{3}{|c|} {Angular} \\
				& \multicolumn{3}{|c|} {(NRMSE [\%])} & \multicolumn{3}{|c|} {gain [X]} & \multicolumn{3}{|c|} {(NRMSE [\%])} & \multicolumn{3}{|c|} {(NRMSE [\%]) }  & \multicolumn{3}{|c|} {error [deg.]} & \multicolumn{3}{|c|} {precision  } \\
				
				\cline{2-19}
				
				&Brain &WM & GM  &  \multicolumn{3}{|l|} {Brain} & Brain & WM & GM & Brain & WM & GM  & \multicolumn{3}{|l|} {WM} & \multicolumn{3}{|l|} {WM}  \\
				\hline 
				
				\multicolumn{1}{|l|}{Original-660${\mu}m$} & 32 & 34 & 29 &  \multicolumn{3}{|l|} {1} & 43 & 29& 61  & 14  & 8 & 16  & \multicolumn{3}{|l|} {15.7} & \multicolumn{3}{|l|} {0.709}  \\
				\hline 
				\multicolumn{1}{|l|}{MPPCA-660${\mu}m$}  & 28 & 29 & 25  &  \multicolumn{3}{|l|} {1.63 } & 32 & 23  & 40 & 9  & 6 & 10 & \multicolumn{3}{|l|} {14.3} & \multicolumn{3}{|l|} {0.757}  \\
				\hline 
				\multicolumn{1}{|l|}{KPCA-660${\mu}m$}  & 22 & 23& 20&  \multicolumn{3}{|l|} {2.25} & 27  & 21 & 30& 8 & 5 & 9 & \multicolumn{3}{|l|} {13.1} & \multicolumn{3}{|l|} {0.787}  \\
				\hline 
				
				
				\multicolumn{1}{|l|}{Original-860${\mu}m$}  & 52 & 57 & 48  &  \multicolumn{3}{|l|} {1} & 51 & 39 & 65 & 21 & 26 & 14   & \multicolumn{3}{|l|} {15.8} & \multicolumn{3}{|l|} {0.705}  \\
				\hline 
				
				
				\multicolumn{1}{|l|}{MPPCA-860${\mu}m$}  & 40& 42  & 36  &  \multicolumn{3}{|l|} {1.81} & 38  & 32  & 44  & 15  & 20  & 9  & \multicolumn{3}{|l|} {13.7} & \multicolumn{3}{|l|} {0.705}  \\
				\hline 
				
				\multicolumn{1}{|l|}{KPCA-860${\mu}m$} & 32 & 34   &33  &  \multicolumn{3}{|l|} {2.71} & 32 & 29 & 35  & 14  & 19  & 8  & \multicolumn{3}{|l|} {12.6} & \multicolumn{3}{|l|} {0.731}  \\
				\hline

		\end{tabular}
	\end{center}
	\caption{ Quantitative results from experiment with in-vivo  human brain submillimeter resolution dMRI data. Note that in all cases i.e., brain, white matter (WM), and gray matter (GM),  KPCA denoising achieves better results than MPPCA. }\label{tab:table_new}
\end{table*}

\begin{figure*}[h]
	\centering
	\includegraphics{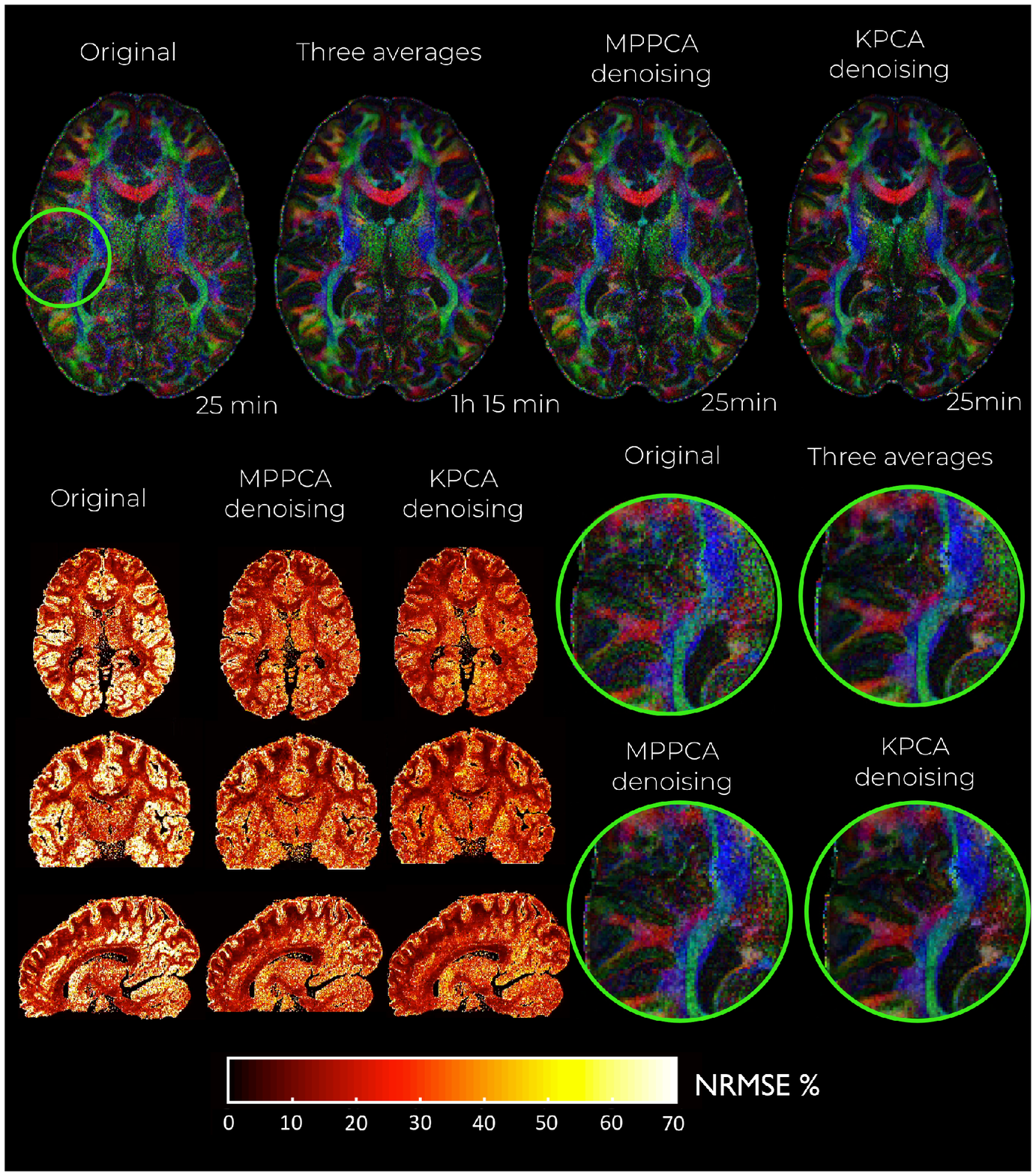} 
	\caption{Color-encoded FA maps of the denoised DW images at 660 ${\mu}m$ isotropic resolution and b-value of $b $ = 1500 $s/{\text{mm}}^2$ as well as corresponding NRMSE maps. Note the better delineation of cortical gray matter in KPCA denoising.}\label{fig:FA660}
\end{figure*}

In good agreement with the findings from the simulation experiment, denoising improves diffusion parameter estimation, and in particular, KPCA denoising helps  estimate quantitative parameters with lower statistical error.  NRMSE of both estimated FA and MD are considerably lower (Table \ref{tab:table_new}) when DTI is applied after KPCA denoising compared to MPPCA. The improvement is significantly noticeable in gray matter.  Cortical gray matter seems better delineated in the FA maps that are obtained after denoising data with KPCA (see color-encoded FA maps in Fig. \ref{fig:FA660}).  This is highly relevant since mapping cortical gray matter is one of the main motivations of ultra-high resolution dMRI protocols \cite{McNab2009}. In fact,  error maps also suggest that the estimation of FA is improved to a greater extent in cortical areas. Color-encoded FA maps as well as errors map for the 860 ${\mu}m$ case are shown in Fig. S4 of the supplementary material. 

\begin{figure*}[h]
	\centering
	\includegraphics{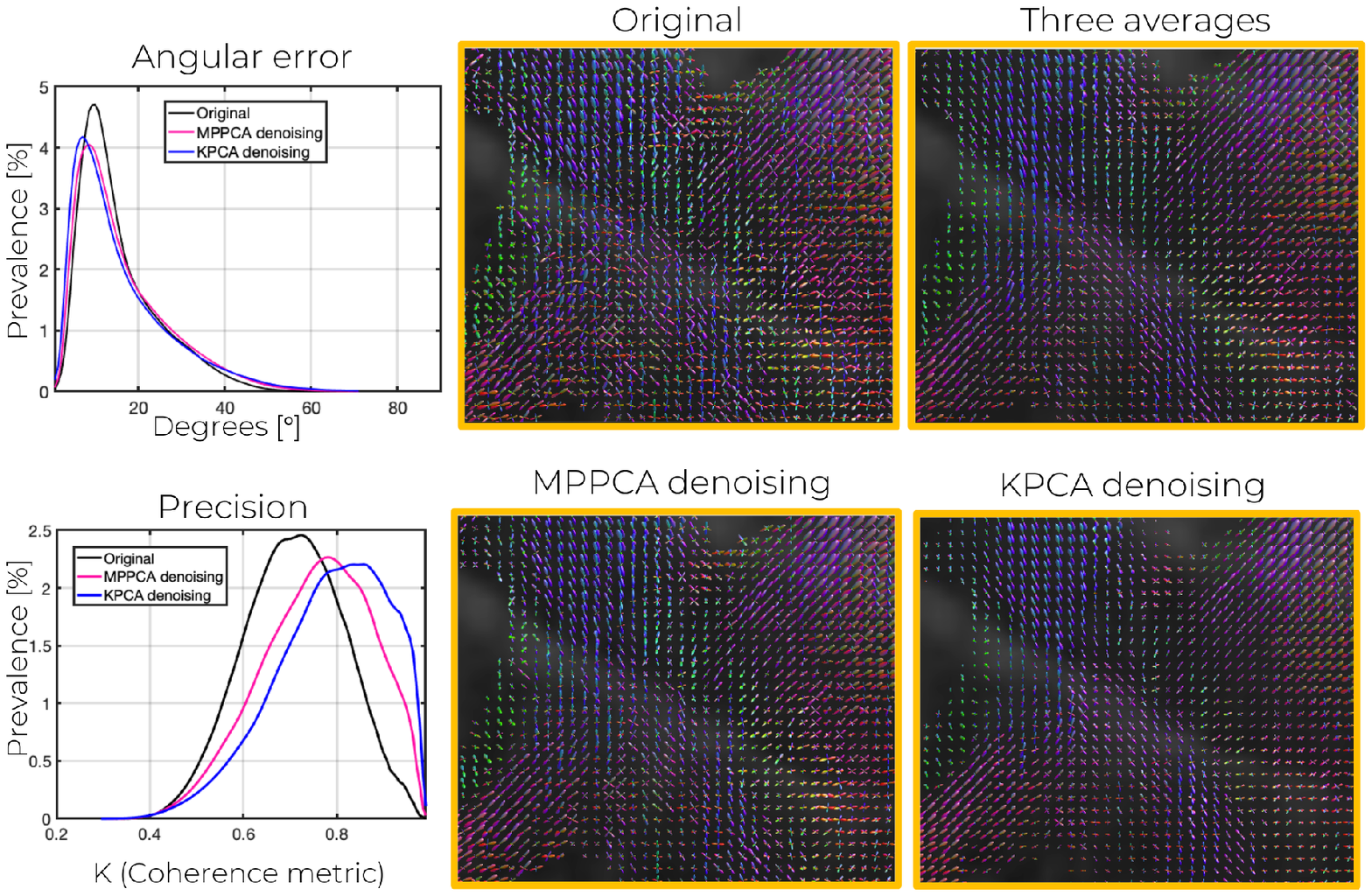} 
	\caption{ Angular error as well as angular precision, probed by coherence metric $\kappa$, for the peaks of the fODFS estimated with CSD after denoising the  660 ${\mu}m$ isotropic resolution DW images  ($b $ = 1500 $s/{\text{mm}}^2$). Further, corresponding fODFs maps in a representative  crossing-fibers area are displayed. Observe the lower variability in the fODFs of  KPCA denoising compared to MPPCA.}\label{fig:ODFs660}
\end{figure*}

fODF estimation becomes more robust after denoising, and more accurate and precise angular directions can be achieved if data is first denoised with KPCA.  As shown in Table \ref{tab:table_new}, lower angular errors (mean of the errors for the first, second, and third peak) are achieved with KPCA. Graphs of the prevalence/probability of angular errors in the 660 ${\mu}m$ data are shown in Fig. \ref{fig:ODFs660} and Fig. S5 (860 ${\mu}m$). 

Clearly, the distribution of the KPCA angular errors is skewed to the left more than that of  MPPCA and the original data, demonstrating lower angular errors in the white matter map obtained from the dataset denoised with KPCA rather than that obtained from the MPPCA denoised or original data. Graphs of the fODFs plotted in the three-fiber crossing area of Fig. \ref{fig:ODFs660} shows lower peak variability with KPCA denoising, a direct consequence of higher noise suppression.   The coherence metric, $\kappa$, proposed in \cite{Jones2003} is in agreement with this observation.  As shown in the plot, both in Fig. \ref{fig:ODFs660} and Fig. S5, the prevalence graphs of $\kappa$ for the KPCA are more right skewed  than that of MPPCA or the original data.  As a result, overall coherence metric values are higher for KPCA (Table \ref{tab:table_new}), indicating higher angular precision could be achieved if data is denoised first with KPCA.

\subsection{Capturing non-linear coil and diffusion redundancy simultaneously} 
\begin{figure*}[h]
	\centering
	\includegraphics{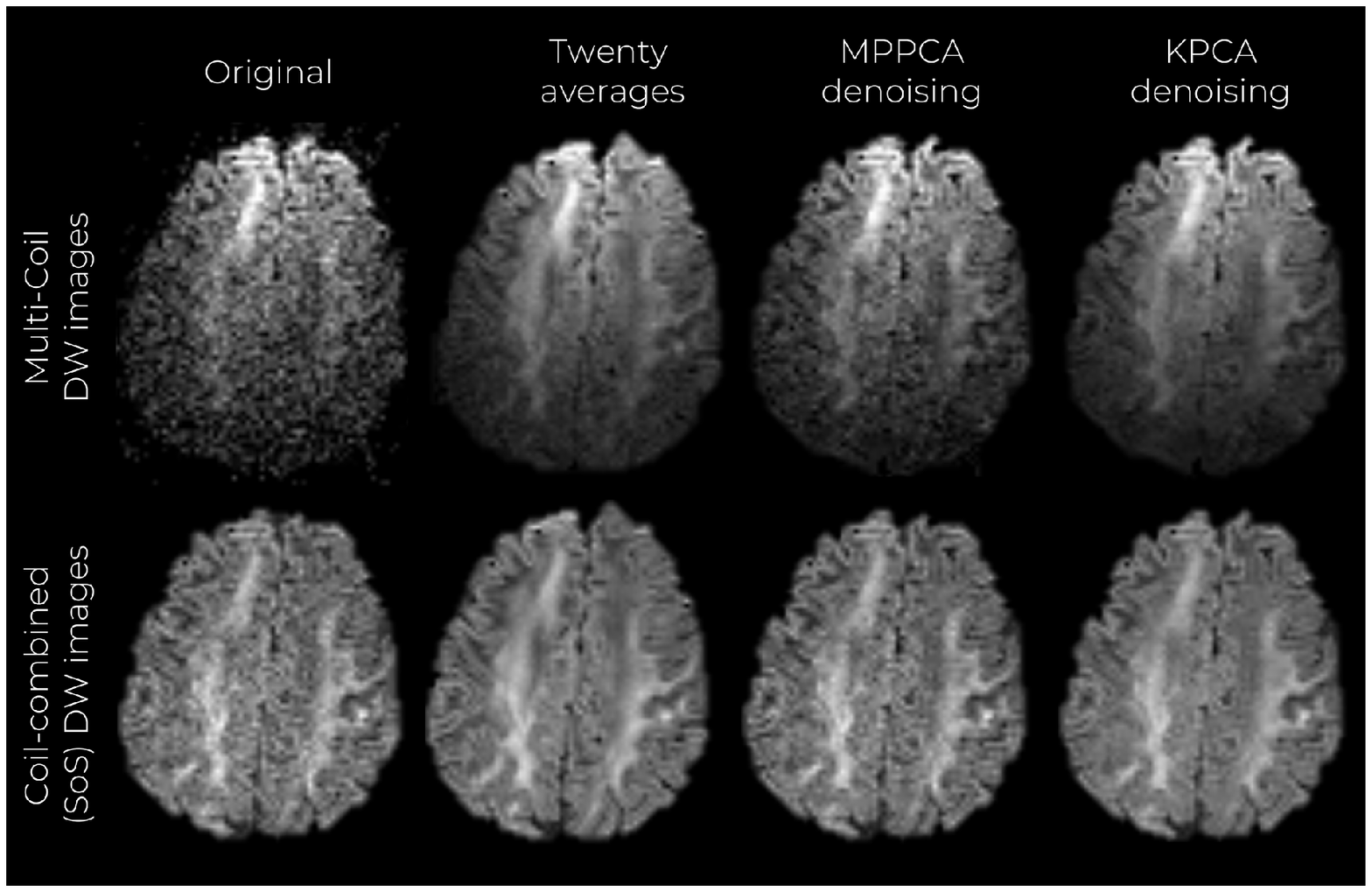} 
	\caption{Mid-axial, coronal and sagittal slices of  multi-coil denoised DW images. DW images coil-combined with the SoS method are also shown. }\label{fig:DWIcoil}
\end{figure*}

Coil DWI images as well as  coil-combined DW images are presented in Fig. \ref{fig:DWIcoil}. The Sum of Squares (SoS) method was employed for coil combination. 

As in the previous experiments, KPCA achieves higher noise suppression than MPPCA, and the result is comparable to the twenty average case. As expected, differences in noise reduction are less notorious in the SoS images, as this technique already denoises the data due to averaging. However, higher noise reduction and good structure preservation is still observed by inspecting the DW images that are denoised with KPCA. Similar to the experiment with high-resolution dMRI data, $\sigma$-normalized residual maps of the multi-coil data show no anatomical features, and signal preservation is confirmed by statistical residual analysis (Fig. S6 of the supplementary material). Furthermore, normalized residuals follow a Gaussian distribution with standard deviation less than one. Noise maps presented in Fig. S7 show higher SNR enhancement when KPCA denoising is applied, compared to MPPCA, i.e., 2.48 X and 1.73 X, respectively. NRMSE values (maps in Fig. S8) in the whole brain were also lower for KPCA than MPPCA. Improved estimation of FA and MD compared to MPPCA is achieved as well (see Table S4. in the supplementary file).

\section{Discussion}
We have shown using realistic simulations and in-vivo dMRI experiments that it is possible to achieve superior noise suppression than state-of-the-art linear PCA denoising while preserving the dMRI image structure, if non-linear redundancies in the data are exploited. No signal structure is removed, as confirmed by the residual analysis. The KPCA denoising methodology can be used to enhance the typically low SNR of dMRI protocols, without compromising signal integrity.  

To exploit non-linear redundancy of the dMRI signal the key point is to apply PCA in a reproducing Kernel Hilbert space where the low-rank assumption of the covariance matrix holds to a greater extent than in the canonical linear PCA space. As the kernel determines the feature space, the choice of the kernel is an interesting problem that deserves to be discussed. We motivated the selection of the Gaussian kernel in Section \ref{ss:kernel}. Though it provided excellent results, other kernels that are specifically tailored to certain features of the dMRI signal, for example, the angular information can be used. This could be accomplished by defining a corresponding spherical covariance function for the diffusion directions, as done in \cite{Andersson2015,Wu2019}, and incorporating this covariance matrix into the conventional Gaussian kernel.

The selection of the rank $K_\mathcal{F}$ and the kernel parameters clearly affects the performance. The SURE method allows us to rely on the statistical distribution model of the data, providing the optimal representation in the MSE sense. Originally, the SURE approach was conceived for additive noise models where the covariance of the noise is diagonal and parameterized  by a single noise level $\sigma$. This case gives excellent results in all of our experiments. However, it can be extended to other statistical distributions\cite{Eldar2008}, including Gaussian noise models with arbitrarily complex covariance matrices. This could be of interest for denoising multi-coil data where correlation between channels do exist, and different noise levels can be found for different channels, very often happening with parallel imaging reconstructions methods. Note that in the experiment with in-vivo multi-coil data, no under-sampling was carried out, and hence no statistical correlation existed in our dataset. Finally, it is worth noting that random matrix theory in kernel matrices is of much less interest for optimal rank selection than in the conventional PCA case. The difficulty of tracking noise statistics over the kernel transformation, and the asymptotic approximations necessary to obtain meaningful theoretical results \cite{ElKaroui2010} makes this line of action impractical. This is one of the main benefits of the SURE method. Indeed, though KPCA denoising performs the low-rank decomposition in the feature transformed space, where noise statistics are difficult to model, the optimal selection of the rank and the scale parameter is done in the native space after reconstruction, where the noise distribution model is well defined.

We would like to emphasize the broad applicability of KPCA denoising  beyond conventional dMRI pulse sequences.  We envisage even further benefits of using KPCA denoising compared to conventional PCA in situations where the complexity of the dMRI signal increases. New developments in diffusion sequences such as multidimensional dMRI \cite{Westin2014,Westin2016, Topgaard2017,Topgaard2020} are highly attractive applications for  KPCA denoising. It is part of our future work to extensively evaluate KPCA denoising in tensor-encoding dMRI data \cite{Szczepankiewicz2019} and extend our preliminary experiments on this kind of data. Combination of relaxometry and diffusion MRI data may be another application where the non-trivial redundancy between different modalities could be better exploited with KPCA \cite{Kim2017,Ning2019,Lampinen2020,Grussu2020}.

\section{Conclusion}
We introduce to the dMRI community a novel denoising technique, Kernel PCA,  which goes beyond the linear compressibility assumption of PCA-based methods and exploits the non-linear redundancies that is intrinsic to dMRI data. Substantially superior SNR-enhanced dMRI data can be obtained compared to PCA, without compromising signal integrity, in a short-computation time, and with no manual parameter tunning. We showcase the power of KPCA denoising with several in-vivo whole human brain submillimeter resolution datasets as well as conventional spatial resolution multi-coil dMRI data. Improved diffusion parameter estimation was observed in all cases compared to state-of-the-art PCA denoising, e.g., Marchenko-Pastur PCA (MPPCA).  We believe KPCA denoising could be beneficial in any diffusion MRI processing pipeline and particularly critical when processing very low SNR data, as in high-resolution dMRI.  

\section*{Acknowledgments}

We acknowledge funding support from the following National Institute of Health (NIH) grant: R01MH116173 (PIs: Setsompop, Rathi).


\end{document}